\documentclass[]{interact}
 \pdfoutput=1 
\usepackage{xspace}
\usepackage{epstopdf}
\usepackage[caption=false]{subfig}
\usepackage[dvipsnames]{xcolor}
\usepackage{xcolor,soul}

\usepackage[numbers,sort&compress,merge]{natbib}
\bibpunct[, ]{[}{]}{,}{n}{,}{,}
\renewcommand\bibfont{\fontsize{10}{12}\selectfont}

\theoremstyle{plain}

\theoremstyle{definition}

\theoremstyle{remark}

\newcommand{\HC}{C$_2$H$_2^+$\xspace}

\newcommand{\wn}{cm$^{-1}$\xspace}

\begin{document}

\articletype{Dieter Gerlich Special Issue}

\title{High-resolution spectroscopy of the $\nu_3$ antisymmetric C-H stretch of \HC
using leak-out action spectroscopy}

\author{
\name{Stephan Schlemmer\textsuperscript{a} and Eline Plaar\textsuperscript{a} and Divita Gupta\textsuperscript{a} and Weslley Guilherme Dias de Paiva Silva\textsuperscript{a} and Thomas Salomon \textsuperscript{a} 
and  Oskar Asvany\textsuperscript{a}\thanks{CONTACT O. Asvany. Email: asvany@ph1.uni-koeln.de}}
\affil{\textsuperscript{a}
I.  Physikalisches Institut, Universit\"at zu K\"oln, Z\"ulpicher Str.~77, 50937 K\"oln, Germany}
}

\maketitle

\begin{abstract}
The antisymmetric C-H stretching vibration $\nu_3$ ($^2\Pi$ $\leftarrow$ $^2\Pi$) of
ionized acetylene, \HC, has been revisited using a cryogenic 22-pole ion trap machine. 
Two action spectroscopic techniques,  the novel  leak-out spectroscopy (LOS) method and the 
more established laser-induced reactions  (LIR) method, 
are applied and compared.
Mass selectivity and cryogenic temperatures down to 4~K enabled 
the observation of uncontaminated spectra in which the $\Lambda$-doubling components 
of this open-shell molecule are mostly well resolved, leading to
a slight refinement of the spectroscopic parameters. 
\end{abstract}



\section{Introduction}

Given the importance of molecular cations to astrochemistry, these ions have been investigated early on with high-resolution 
rovibrational spectroscopy. Using a cooled discharge tube in conjunction
with a tunable difference-frequency mixing laser source, many 
of these important species, in particular carbocations, were probed for the first time
in the group of Takeshi Oka in Chicago~\cite{oka80,cro88, tan99, whi99}. One of these fundamental ions 
is cationic acetylene, \HC ($\tilde{X}^2\Pi_u$), for which the  antisymmetric C-H stretch 
$\nu_3$ ($^2\Pi_u$ $\leftarrow$ $^2\Pi_u$) has been recorded in rotational resolution~\cite{cro87,jag92}.
A strong perturbation for higher rotational levels of  the excited state was observed~\cite{jag92}, 
requiring the application of ground state combination differences as an analytical tool.
Thus the spectroscopic parameters of \HC, as well as those of
its deuterated siblings HCCD$^+$ and C$_2$D$_2^+$~\cite{roe91,jag92} were obtained
for the first time.


Due to the possibility of cryogenic cooling and mass selectivity, it became 
apparent in the 1990s that action spectroscopy performed in  multipole 
ion traps has several advantages over conventional spectroscopy. 
It was in particular Dieter Gerlich, who pushed  forward with the development of 
his 22-pole ion trap~\cite{ger95}. This enabled cooling of trapped molecular ions for ion chemistry~\cite{ger92a} 
and spectroscopic applications, first in Freiburg, and later in Chemnitz. 
Together with  Stephan Schlemmer, they developed the action 
spectroscopy method laser induced reactions  (LIR), 
and first applied it to the electronic spectroscopy of N$_2^+$ ~\cite{scl99}. 
A few years later,  exploiting the endothermic reaction \HC + H$_2$ + h$\nu$ $\rightarrow$
C$_2$H$_3^+$ + H, LIR  was also proven to work well for vibrational spectroscopy,
as demonstrated for the  $\nu_3$ band of \HC \cite{scl02}. Because of the large bandwidth
of the applied tunable IR laser ($\Delta\nu=0.24$~\wn), and also due to 
the relatively high trap temperature of $T=90$~K, the $\Lambda$-doubling and the 
rotational details of the Q-branch were unfortunately not resolved (see Fig.~1 in \cite{scl02}).
Nevertheless,  LIR laid the ground for the first direct (but low-resolution) investigation of the 
cis-bending vibration $\nu_5$ of this molecule~\cite{scl05a,asv05},
which is now known in high-resolution by ZEKE-spectroscopy \cite{yan06,tan06}.

In this  paper, we re-investigate the  $\nu_3$ band of \HC using action spectroscopic methods,  but  this time 
applying cryogenic buffer gas cooling to the lowest possible temperatures and using  high-resolution light sources.
Apart from exploiting LIR,  we apply an action spectroscopy method called leak-out-spectroscopy  (LOS),
recently developed in the Cologne laboratories~\cite{scm22a}. As LOS does not require a reaction partner,
but only an inert collision partner (typically a noble gas atom), 
it is more generally applicable than LIR.
In addition, LOS is very effective and can be applied down to a trap temperature of 4~K.
We compare the spectra obtained by both action spectroscopic methods, 
and refine the spectroscopic parameters of the $\nu_3$ band first reported 40 years ago.


\section{Experimental}

The rovibrational transitions of \HC were measured in  COLTRAP, a 4~K 22-pole ion trap instrument 
which has been described in detail previously \cite{asv10,asv14}. 
Ions were generated in a storage ion source by electron impact ionization ($E_e = 32.5$~eV) of 
a 50:50 acetylene (C$_{2}$H$_{2}$; \textit{Messer} $\geq$ 99.6\% purity) - helium (\textit{Linde} $\geq$99.999\% purity) gas mixture.   
The initially produced ions are pulsed into a quadrupole mass analyzer
and  selected for mass $m=26$~u.
The remaining  ions (on the order of several ten thousands)
are guided into a 22-pole ion trap mounted on a 4~K cold head. 
Efficient trapping and thermalization of the ions is achieved through collisions 
with Helium buffer gas which is continuously introduced at low  densities.
Additionally, in order to provide a proper collisional partner for the novel LOS approach and 
to avoid extensive freeze-out at the trap surfaces, a 1:3 mixture of Neon diluted 
in Helium is pulsed into the trap through a piezoelectrically actuated valve at 
the beginning of each trapping cycle.

\begin{figure}[t!]
\centering
\includegraphics[width=1\columnwidth]{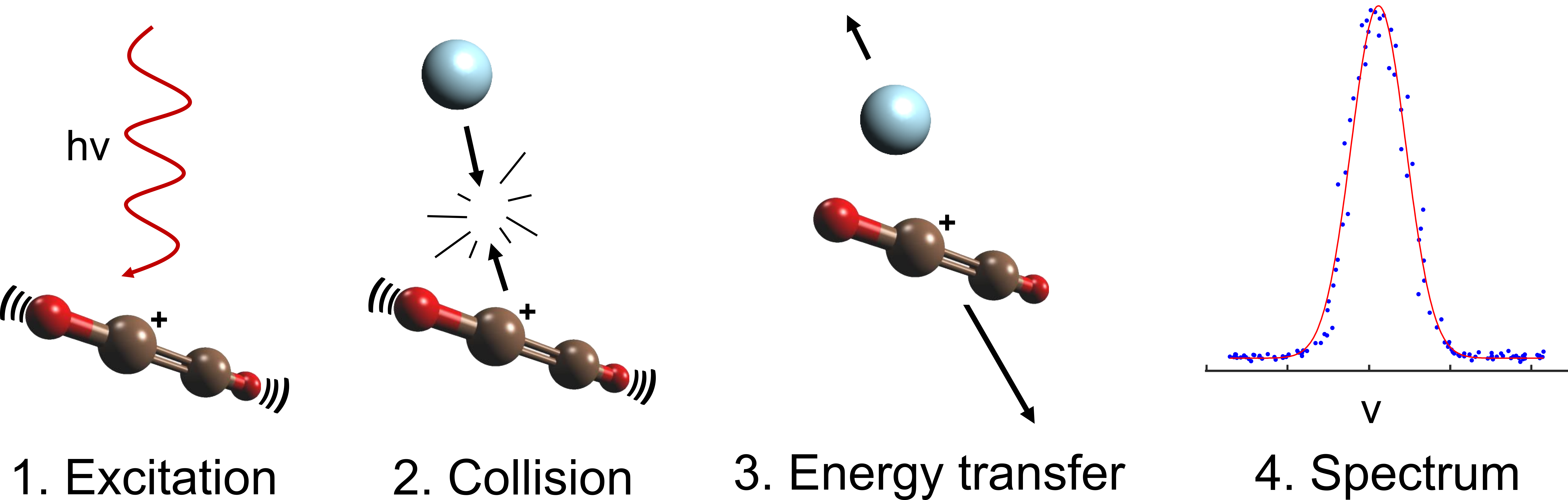}
\caption{\label{fig:LOS} 
Principle of leak-out-spectroscopy (LOS), shown here for \HC in collision with Ne atoms: 
(1) Several ten thousand \HC ions are held in a 4~K ion trap, and resonantly excited into a vibrationally 
excited state by  laser photons. (2) Neon atoms are co-added to the trap volume and collide with the vibrationally excited ions. (3) After the collision, the vibration-to-translation 
energy transfer (V-T-process) accelerates both collision partners.
(4) The accelerated  ions that overcome the trap barrier are guided into the detector and 
are counted as a function of the laser wavenumber to produce the rovibrational spectrum.}
\end{figure}

To detect the rovibrational transitions of \HC, 
we used two different action spectroscopy methods, one of them being
the recently developed and very sensitive LOS technique. 
The LOS method is described in detail in a recent publication~\cite{scm22a} and  
 illustrated here in Fig.~\ref{fig:LOS}.
LOS is based on the escape of a trapped ion after collision-induced transfer of its internal energy to kinetic energy.  
During the  trapping time, the ion ensemble is irradiated for 100~ms by an IR beam traversing  the trap. 
If the  radiation frequency is resonant with a rovibrational transition of \HC, 
the  ion is excited into the $\nu_3$ state.
A vibrationally inelastic collision 
between the excited \HC ion and a Neon atom then leads to a redistribution of the 
internal excitation and partly to kinetic energy 
of both collision partners. 
Then,  the accelerated \HC ion has enough energy to pass the exit potential 
barrier (of about 100 mV) which keeps the other, thermalized ions trapped, 
and to fly out towards the ion detector. The trapping cycle is repeated at a frequency of 1~Hz 
and the escaping \HC ions are counted as a function of the laser wavenumber to record the rovibrational spectrum.

In  former measurement campaigns, we used LIR as the action spectroscopic method,
described in detail in references \cite{scl02,scl05a,asv05} for the case of \HC. 
In brief, the slightly endothermic reaction given in the introduction is exploited, and the enhancement of 
the reaction product C$_2$H$_3^+$ signals the absorption of a photon by \HC. 
The  number of product ions C$_2$H$_3^+$ is thus detected as a function of 
the laser wavenumber to record the rovibrational spectrum.
Typically, the neutral reaction partner H$_2$ is continuously added to the trap leading to a constant number density. In order to avoid freeze-out of H$_2$, the trap is operated 
at slightly elevated temperatures above 10~K.

In both action spectroscopy campaigns, we used continuous-wave (cw) high-resolution  optical 
parametric oscillators (OPOs),  operating in the 3~$\mu$m spectral region, as IR light sources.
For the LOS campaign, we used a commercial Aculight Argos Model 2400 C. 
The IR beam entered the vacuum environment via a 0.6~mm thick diamond window (Diamond Materials GmbH), traversed the 22-pole trap, and exited the vacuum system via a CaF window after which it was absorbed by a power meter.
The power was on the order of 200~mW. 
The irradiation time was controlled by a laser shutter (Thorlabs model SH05).
The frequency of the IR radiation has been measured by a wavemeter (Bristol model 621A) 
with an accuracy of  0.001~cm$^{-1}$ (in case it is well adjusted).
We did additional calibration measurements with 
neutral C$_2$H$_2$ contained in an absorption cell,
whose exact frequencies are given in the HITRAN database \cite{gor22}.
After the calibration, we shifted our data up by 0.007~cm$^{-1}$. 
With this procedure, the accuracy of the data is expected to be on the order of 0.001~cm$^{-1}$.


\section{Rovibrational spectrum obtained with LOS}

 \HC is a linear  open-shell molecule with ground state   $\tilde{X}^2\Pi_u$. 
 The coupling of the electron spin ($S=1/2$) and 
 orbital ($\Lambda=1$) angular momenta leads to two spin-orbit 
 fine structure components, the lower-lying $^2\Pi_{3/2}$ (F$_1$) and  
 the $^2\Pi_{1/2}$ (F$_2$) manifold, which is about 30~\wn higher in energy.
 In addition, the physical rotation of the molecule lifts the degeneracy of 
  the orbital angular momentum,  leading to $\Lambda$-doubling, 
  which is resolvable in our measurements for higher rotational quantum numbers.

\begin{figure} [h!]
\centering
\includegraphics[width=1.0\columnwidth]{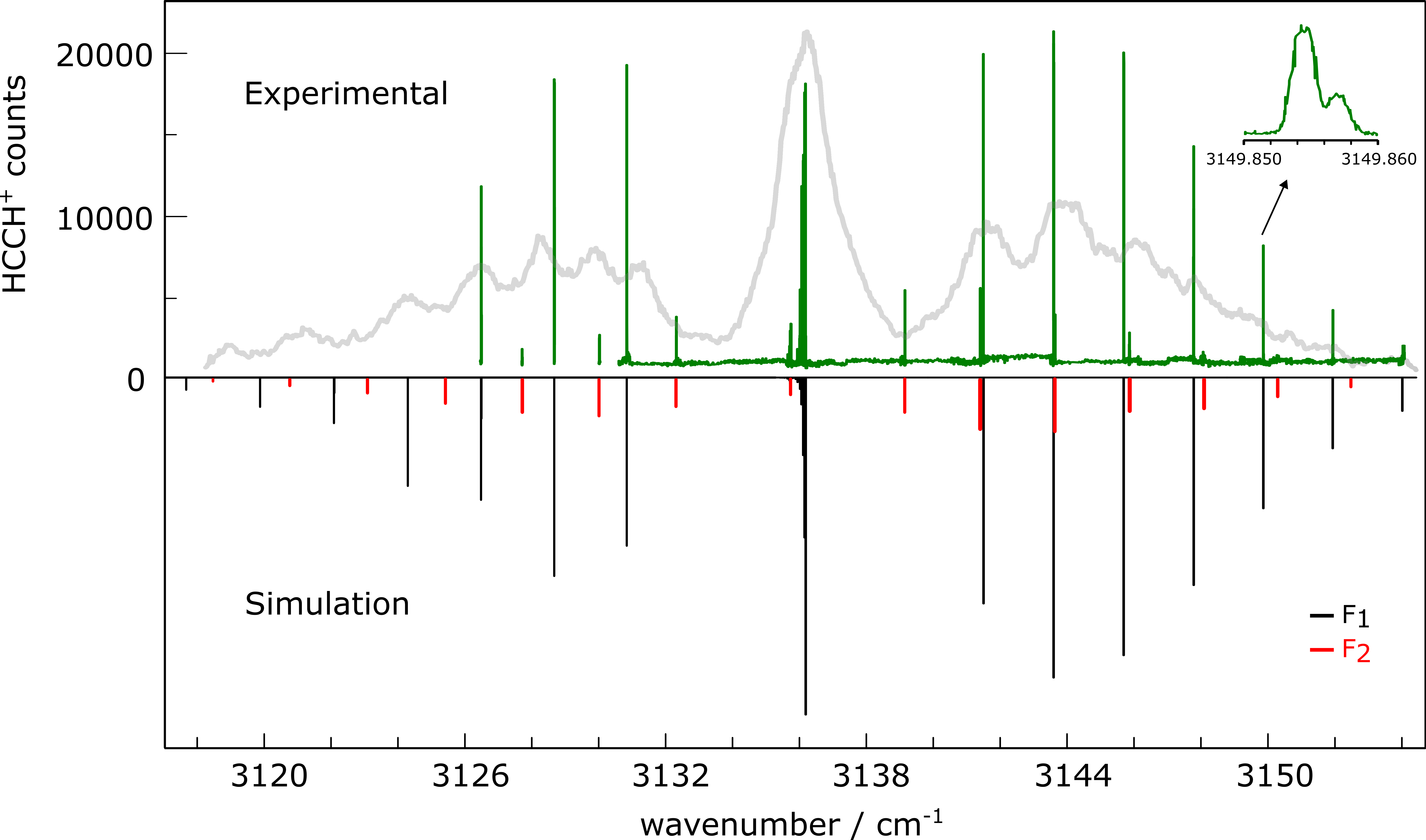}
\caption{\label{fig:spectrum} Rovibrational spectrum of the $\nu_3$ antisymmetric C-H stretch band of \HC,  recorded using the novel leak-out spectroscopy (LOS) method. The upper part is the measured high-resolution spectrum (in green), while the bottom part is a simulation based on the fitted parameters given in Table~\ref{tab:fit} at 26~K. In the simulation, the lower-energy F$_1$ and the higher-energy F$_2$ components are color-coded in black and red, respectively. The inset shows the F$_1$ component of the J' $\leftarrow$ J" = 6.5 $\leftarrow$ 5.5 transition in which the $\Lambda$-type doubling (e and f) is resolved. 
The grey trace in the upper part shows an overview spectrum also obtained by LOS with a low-resolution pulsed OPO (Laser Vision).
The green trace is provided as a separate data file in the supplementary material.}
\end{figure}

 The overview LOS-spectrum of \HC is shown in Fig.~\ref{fig:spectrum}, together 
 with a spectroscopic simulation performed with PGOPHER~\cite{wes17}. 
 The measured Q-branch is shown in more detail in Fig.~\ref{fig:comp}.
 Unfortunately, because of the limited tuning range of the OPO module, only part of the 
 P-branch could be recorded. As expected, 
the lower-lying F$_1$ component is much stronger than   F$_2$ at our
cryogenic conditions, as seen by the color-coded simulation in Fig.~\ref{fig:spectrum}.
The inset in Fig.~\ref{fig:spectrum} shows the 
F$_1$ line $J' \leftarrow J" = 6.5 \leftarrow 5.5$,
with resolved $\Lambda$-doubling (separated by $\approx$0.0025~\wn), exhibiting the 
expected 3:1 ratio between the two ortho-para components.
By analyzing such resolved lines, in particular those in the Q-branch, 
a kinetic ion temperature of
about $T_{kin}\approx20$~K is obtained, while the simulation in  Fig.~\ref{fig:spectrum}
gives an estimation of the rotational temperature of about $T_{rot}\approx22 - 26$~K.

The list of $\nu_3$-lines measured with LOS is given in  
 the supplementary material. There, the lines measured with LOS 
are compared to those measured by the Oka group~\cite{jag92},
showing that our cryogenic cooling permits to resolve
many $\Lambda$-doublets for the first time (see also Fig.~\ref{fig:comp}).
On average, the lines from the Oka measurements are only about 0.0007~\wn lower than our LOS data, 
indicating a  good calibration of both data sets.

\section{LOS versus LIR}

\begin{figure}[b!]
\centering
\includegraphics[width=1.0\columnwidth]{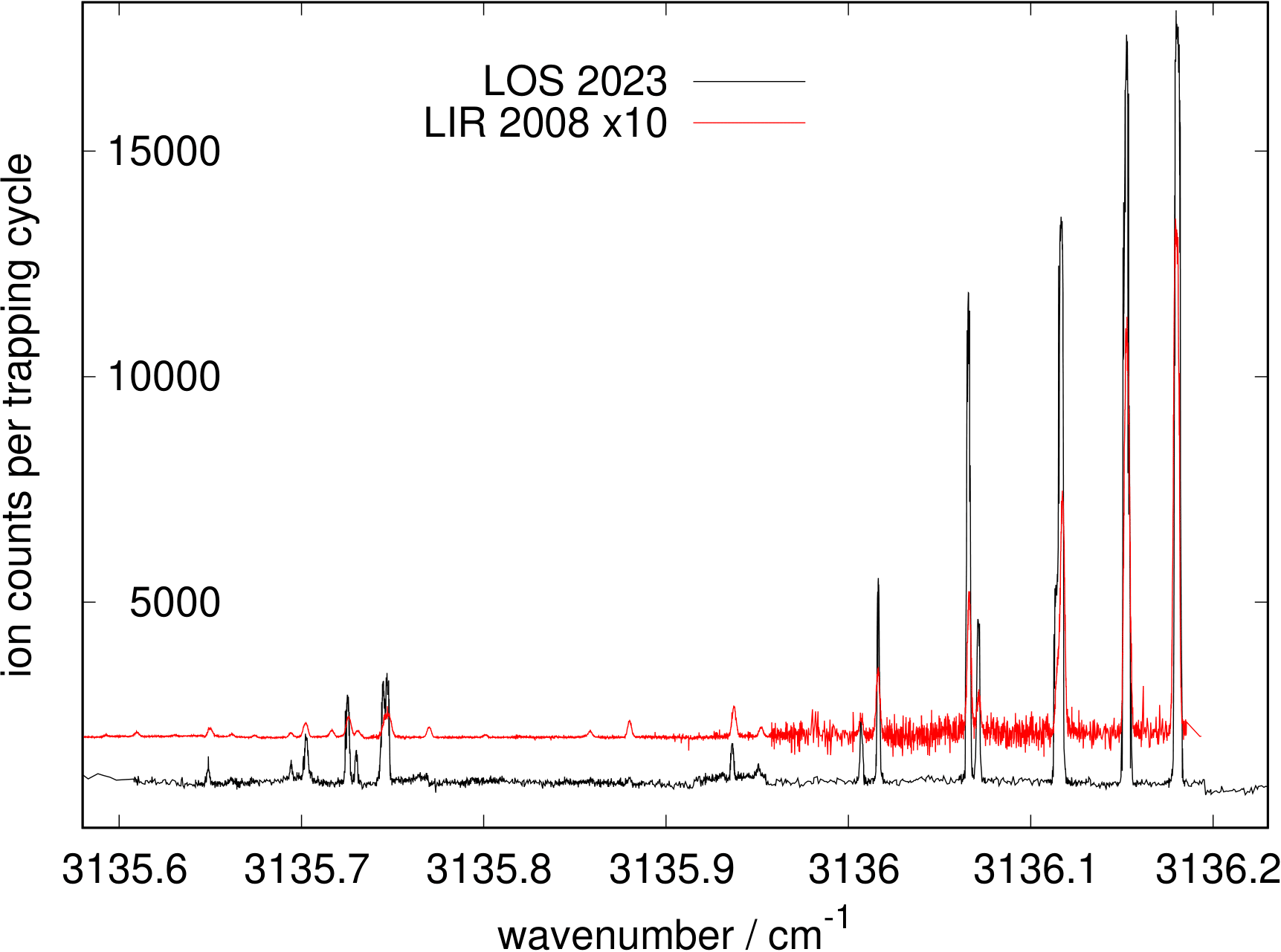}
\caption{\label{fig:comp} 
Comparison of measurements of the $\nu_3$ Q-branch of \HC using the methods LOS and LIR.
With LIR, the laser traversing the trap is promoting the reaction 
\HC + H$_2$ $\rightarrow$ C$_2$H$_3^+$ + H at about $T$~=~13.5~K, and the C$_2$H$_3^+$ products
created during the trapping time are counted as a function of the laser wavenumber.
In this Figure, the product counts have been multiplied by a factor of 10
for better comparison. 
For LOS, the trap is held at a nominal temperature of 4~K, 
and one simply counts the \HC ions which are kicked out of the 
ion trap after the resonantly excited ions collide with Ne atoms.  
Higher ion counts have been reached for LOS and therefore signal-to-noise ratio achieved in the LOS measurements is better than for LIR.}
\end{figure}

Fig.~\ref{fig:comp} shows a zoom of the Q-branch region comparing  the measurements
done with LOS and LIR. Please observe that in the LIR 
measurements the action signal is given by the ion counts of the 
accumulated reaction products C$_2$H$_3^+$, measured in a small time window of 
a few ms {\it after} the irradiation/trapping cycle,
while for LOS the targeted ions \HC are directly detected {\it during} 
 the irradiation/trapping cycle, lasting about 100~ms.
In the Cologne group, two measurement campaigns have been performed with LIR to 
investigate \HC, one in 2008 and another in 2014 (both unpublished). 
The LIR-measurement presented in  Fig.~\ref{fig:comp}
is the one done in 2008, performed 
in a different ion trap machine (a 10~K trap machine called LIRTRAP, see e.g.~\cite{scl05a}), and using a home-built OPO.
The  nominal ion trap temperature was about 13.5~K.
It should be noted here that LIR can also operate at temperatures below the condensation temperature of the reaction gas (H$_2$ in this case), 
by simply pulsing the reaction gas into the trap as done in later works \cite{asv12,asv15}. 
However, this was not implemented in the LIR measurement shown in Fig.~\ref{fig:comp}.

Both action methods detect the two fine-structure components F1 and F2,
and nicely resolve the $\Lambda$-doubling for lines 
with J' $\leftarrow$ J" = 4.5 $\leftarrow$ 4.5 and higher in F1.
The LOS measurement exhibits a somewhat lower kinetic
temperature ($T_{kin}\approx20$~K or even lower) 
compared to the LIR measurement  ($T_{kin}\approx27$~K), 
as determined from the Doppler widths of well-separated lines.
The discrepancy to the nominal temperatures ($T=4$~K and $T=13.5$~K, respectively)
is assumed to be caused by heating effects when the ions undergo collisions in the RF field of the ion trap~\cite{asv09}.
This can be particularly severe by the use of the heavy collision partner Neon in the case of LOS, 
while LIR uses only the light collision partners He and H$_2$.
Also, the rotational temperature obtained with LOS is somewhat lower, on the order of $T_{rot}\approx22$~K,
as determined from the fading of the R-branch transitions in Fig~\ref{fig:spectrum}.
In summary,  the very good signal-to-noise ratio 
renders LOS the preferred action spectroscopic method in the case of \HC.
The overall advantage of LOS, however, is that it can be generally applied to any cation or anion, 
without the need of a suitable low-temperature reaction.


\section{Spectroscopic constants}

The assigned rovibrational transitions observed with our high-resolution LOS method  were fitted 
using a $^2\Pi$ $\leftarrow$ $^2\Pi$ linear top Hamiltonian as implemented in 
Western's PGOPHER program \cite{wes17} to determine the spectroscopic parameters of 
\HC for the ground  and the $\nu_3$~=~1 vibrational states. 
The determined parameters comprise the rotational constants ($B$), the spin-orbit coupling constants ($A$), $\Lambda$-doubling constants ($q,p$), the quartic centrifugal distortion constants ($D$) of the ground and vibrationally excited states, as well as the $\nu_3$ band origin. 

As already pointed out by Jagod et al~\cite{jag92}, the  $\nu_3$ 
vibrationally excited state is heavily perturbed,
most likely by Renner-Teller states involving the $\nu_4$ and $\nu_5$ bending vibrations, thus, making a 
fit of the unperturbed ground state via combination differences mandatory.
We followed this recipe. 
To better determine the distortion constant $D_0$ and the $\Lambda$-doubling constants, 
we included 82 lines from Oka with higher rotational quantum numbers where the  $\Lambda$-doubling has been resolved.
These were the lines with frequencies below 3100~\wn (P-branch) and above 3158~\wn (R-branch)  \cite{jag92}.
The fit, with 68 combination differences,  has a reasonably good quality with an rms of obs-calc values of 0.0022~\wn 
(more than half of the used lines are from Jagod et al~\cite{jag92} with an uncertainty of 0.003~\wn) and a corresponding 
weighted rms of 0.71.
The derived spectroscopic parameters from this 
ground state combination difference fit are given 
in Table~\ref{tab:fit}. All used transition frequencies and fit residuals are 
listed in the supporting information material. 
The fit results from Jagod et al~\cite{jag92} 
are very close to ours and are only marginally improved by our data;
they are repeated in Table~\ref{tab:fit}.

\begin{table} [b!]
\caption{\label{tab:fit} Spectroscopic parameters of \HC. Those of the ground state have been obtained by 
using ground state combination differences of our data and those of Jagod et al~\cite{jag92}. 
All listed values are given in \wn with their respective uncertainties provided in parenthesis.
The results previously reported by Jagod et al~\cite{jag92} are also included in the Table.\\} 
\begin{center}
\begin{tabular}{l|  r@{}l  | r@{}l | r@{}l  | r@{}l}
\hline
           & \multicolumn{4}{c}{this work}    & \multicolumn{4}{c}{Jagod et al~\cite{jag92}}            \\
Parameter  & \multicolumn{2}{c}{ground} &  \multicolumn{2}{c}{$\nu_3$} & \multicolumn{2}{c}{ground}    &  \multicolumn{2}{c}{$\nu_3$}         \\
\hline
$\nu$      &        &                      &  3135.&97376(48)     &&&     3135.&9813(10)                  \\
  $B$      &   1.&104636(16)               &  1.&099521(23)       &   1.&10463(2) &  1.&09843(7)   \\
  $A$  	 &   -30.&871(16)                & -30.&42744(79)        &   -30.&91(2)  & -30.&466(2) \\
  $q$      &  -0.&332(13)$\times10^{-3}$   & -0.&352(65)$\times10^{-3}$  &  -0.&359(16)$\times10^{-3}$   & -0.&359$\times10^{-3}$  \emph{$^a$}\\
  $p$      &  -0.&63(13)$\times10^{-3}$    & -0.&63$\times10^{-3}$   \emph{$^a$} &  &    & &                   \\
  $D$      &  1.&600(19)$\times10^{-6}$    & 1.&600$\times10^{-6}$   \emph{$^a$} &  1.&57(2)$\times10^{-6}$    & 1.&57$\times10^{-6}$  \emph{$^a$}   \\
\hline
\end{tabular}
\end{center}
$\emph{$^a$}$  Fixed to the ground state value
\end{table}

In the second step, the ground state spectroscopic parameters were fixed, and the 
parameters of the $\nu_3$ band were fitted. Due to the mentioned heavy perturbation, occurring at higher 
rotational quantum numbers in both F$_1$ and F$_2$ series, only  
our data have been taken into account.
Also, some of our lines with high obs-calc values were excluded from the fit, which are marked 
in the list given in the supplementary information. 
Our spectroscopic parameters of the $\nu_3$ state are also given in Table~\ref{tab:fit}. 
Due to the improved measurement precision of our novel approach, 
these are refined compared  to the results of Jagod et al~\cite{jag92}.
There are also some obvious shifts in the values
(e.g.\ our band origin is about 0.008~\wn lower), which we attribute 
to the ambiguity when fitting a semi-rigid molecular model
to a perturbed system.


\section{Summary and Conclusion}

LOS is a novel action spectroscopic method~\cite{scm22a,gup23,asv23} which needs further validation.
In this work on \HC,  it is  compared with a more established action spectroscopic method, LIR~\cite{scl02},  
showing very consistent spectral features and a slightly better signal-to-noise ratio.
This is an encouraging result, as spectra of hitherto unexplored molecules, obtained with LOS, 
sometimes exhibited  unidentified  lines (see the example of HCCCH$^+$ below), as well
as hot bands in case the trap is operated at temperatures beyond 4~K.
This work thus demonstrates the reliability of LOS, 
and shows that former observations  of unidentified lines are most probably 
intrinsic to the molecule under 
investigation and not caused by our novel method.

The overall advantage of LOS over other action methods is its independence from 
 e.g.\ chemical reactions or temperature ranges, so that it can  be generally applied to any cation or anion.
Also, ions with higher masses can be probed by LOS, simply by applying heavier non-reactive 
collision partners such as N$_2$, Ar, Kr or Xe. 
The low-temperature operation can be preserved by pulsing in He-mixtures of these gases into the trap.
Investigations along these directions
are currently being conducted in the Cologne laboratories.
Only recently, LOS opened the way to investigate the larger sibling of \HC, 
HCCCH$^+$ (also $\tilde{X}^2\Pi_u$). 
Similar to this work,
we probed the antisymmetric C-H stretching vibration $\nu_3$, predicted 
to be in the 3200~\wn region~\cite{bru19}. 
Although we detected strong and somewhat regular P- and R-branches with a suitable 
rotational constant, as well as some
$\Lambda$-doubled lines with a typical 3:1 ratio, unfortunately, this mode
seems even more heavily perturbed than that of \HC.
Again, the reason for these perturbations is most probably a variety of 
Renner-Teller states involving bending vibrations, of which there are three for HCCCH$^+$.  
The perturbations caused by these, as well as the plethora of unidentified interleaved lines, 
make the analysis of this band very challenging.
For instance, a clear identification of the Q-branch and the two fine-structure 
components (F$_1$ and F$_2$) was not possible yet.
High-level {\it ab initio} calculations of its ground state,
in particular of the spin-orbit splitting constant $A$,  would be of great help.

The rotationally resolved spectrum of the antisymmetric $\nu_3$  C-H stretching vibration of \HC recorded in the laboratory of Dieter Gerlich was the first action spectrum involving a classical chemical reaction. This marked the starting point of a rapidly developing field where many mechanisms have been invented as a possible action to make the spectrum \emph{visible}. Professor Gerlich opened this avenue and invented a number of those enabling techniques. LOS is the latest in this row and revisiting the spectrum of \HC demonstrates the advances of the field w.r.t. sensitivity and spectral resolution. LOS allows to record background-free spectra of mass-selected species of seemingly unlimited choice. Kicking out all ions by \emph{one} particular transition, thus,  addressing, e.g., a specific structural isomer, nuclear spin isomer or just an isobaric species, allows for the analysis and preparation of selected samples for subsequent experiments. For instance, this will enable the study of 
isomer-selected reactions in the future.


\section*{Acknowledgements}

The LIR-measurement of the Q-branch depicted in Fig.~\ref{fig:comp} 
has been conducted in the year 2008 in collaboration with Sabrina Gärtner and Jürgen Krieg.
Additional LIR measurements have been performed by Pavol Jusko in 2014 (not shown in this work).
The current work  has been  supported by an
ERC advanced grant (MissIons: 101020583),
the Deutsche Forschungsgemeinschaft (DFG) via SFB 956 (project ID 184018867), sub-project B2,
and the Ger\"atezentrum "Cologne Center for Terahertz Spectroscopy" (DFG SCHL 341/15-1). W.G.D.P.S. thanks the Alexander von Humboldt foundation for funding through a postdoctoral fellowship.
The authors gratefully acknowledge the work done by the electrical and mechanical workshops of the I.~Physi\-kali\-sches Institut over the last years.


\begin{thebibliography}{26}
\providecommand{\url}[1]{\texttt{#1}}
\providecommand{\urlprefix}{URL }

\bibitem{oka80}
T. Oka,  Phys. Rev. Lett.  \textbf{45}, 531--534 (1980).

\bibitem{cro88}
M.W. Crofton, M. Jagod, B.D. Rehfuss, W.A. Kreiner and T. Oka,  J. Chem. Phys.
  \textbf{88}, 666--678 (1988).

\bibitem{tan99}
J. Tang and T. Oka,  Journal of Molecular Spectroscopy  \textbf{196} (1),
  120--130 (1999).

\bibitem{whi99}
E.T. White, J. Tang and T. Oka,  Science  \textbf{284}, 135--137 (1999).

\bibitem{cro87}
M.W. Crofton, M.F. Jagod, B.D. Rehfuss and T. Oka,  J. Chem. Phys.
  \textbf{86}, 3755--3756 (1987).

\bibitem{jag92}
M.F. Jagod, M. R\"osslein, C.M. Gabrys, B.D. Rehfuss, F. Scappini, M.W. Crofton
  and T. Oka,  J. Chem. Phys.  \textbf{97}, 7111--7123 (1992).

\bibitem{roe91}
M. R\"osslein, M.F. Jagod, C.M. Gabrys and T. Oka,  Astrophys. J.
  \textbf{382}, L51--L53 (1991).

\bibitem{ger95}
D. Gerlich,  Phys. Scr.  \textbf{T59}, 256--263 (1995).

\bibitem{ger92a}
D. Gerlich and S. Horning,  Chem. Rev.  \textbf{92}, 1509--1539 (1992).

\bibitem{scl99}
S. Schlemmer, T. Kuhn, E. Lescop and D. Gerlich,  Int. J. Mass Spectrom.
  \textbf{185}, 589--602 (1999).

\bibitem{scl02}
S. Schlemmer, E. Lescop, J. v.~Richthofen and D. Gerlich,  J. Chem. Phys.
  \textbf{117}, 2068--2075 (2002).

\bibitem{scl05a}
S. Schlemmer, O. Asvany and T. Giesen,  Phys. Chem. Chem. Phys.  \textbf{7},
  1592--1600 (2005).

\bibitem{asv05}
O. Asvany, T. Giesen, B. Redlich and S. Schlemmer,  Phys. Rev. Lett.
  \textbf{94}, 073001 (2005).

\bibitem{yan06}
J. Yang and Y. Mo,  J. Phys. Chem. A  \textbf{110} (38), 11001--11009 (2006).

\bibitem{tan06}
S.J. Tang, Y.C. Chou, J.J.M. Lin and Y.C. Hsu,  J. Chem. Phys.  \textbf{125}
  (13), 133201 (2006).

\bibitem{scm22a}
P.C. Schmid, O. Asvany, T. Salomon, S. Thorwirth and S. Schlemmer,  J. Phys.
  Chem. A  \textbf{126} (43), 8111--8117 (2022), PMID: 36278898.

\bibitem{asv10}
O. Asvany, F. Bielau, D. Moratschke, J. Krause and S. Schlemmer,  Rev. Sci.
  Instr.  \textbf{81}, 076102 (2010).

\bibitem{asv14}
O. Asvany, S. Br\"unken, L. Kluge and S. Schlemmer,  Appl. Phys. B
  \textbf{114} (1-2), 203--211 (2014).

\bibitem{gor22}
I. Gordon, L. Rothman, R. Hargreaves, R. Hashemi, E. Karlovets, F. Skinner, E.
  Conway, C. Hill, R. Kochanov, Y. Tan, P. Wcisło, A. Finenko, K. Nelson, P.
  Bernath, M. Birk, V. Boudon, A. Campargue, K. Chance, A. Coustenis, B.
  Drouin, J. Flaud, R. Gamache, J. Hodges, D. Jacquemart, E. Mlawer, A.
  Nikitin, V. Perevalov, M. Rotger, J. Tennyson, G. Toon, H. Tran, V. Tyuterev,
  E. Adkins, A. Baker, A. Barbe, E. Canè, A. Császár, A. Dudaryonok, O.
  Egorov, A. Fleisher, H. Fleurbaey, A. Foltynowicz, T. Furtenbacher, J.
  Harrison, J. Hartmann, V. Horneman, X. Huang, T. Karman, J. Karns, S. Kassi,
  I. Kleiner, V. Kofman, F. Kwabia–Tchana, N. Lavrentieva, T. Lee, D. Long,
  A. Lukashevskaya, O. Lyulin, V. Makhnev, W. Matt, S. Massie, M. Melosso, S.
  Mikhailenko, D. Mondelain, H. Müller, O. Naumenko, A. Perrin, O. Polyansky,
  E. Raddaoui, P. Raston, Z. Reed, M. Rey, C. Richard, R. Tóbiás, I. Sadiek,
  D. Schwenke, E. Starikova, K. Sung, F. Tamassia, S. Tashkun, J. {Vander
  Auwera}, I. Vasilenko, A. Vigasin, G. Villanueva, B. Vispoel, G. Wagner, A.
  Yachmenev and S. Yurchenko,  J. Quant. Spectrosc. Radiat. Transfer
  \textbf{277}, 107949 (2022).

\bibitem{wes17}
C.M. Western,  J. Quant. Spectros. Rad. Transfer  \textbf{186}, 221 -- 242
  (2017).

\bibitem{asv12}
O. Asvany, J. Krieg and S. Schlemmer,  Rev. Sci. Instr.  \textbf{83}, 093110
  (2012).

\bibitem{asv15}
O. Asvany, K.M.T. Yamada, S. Br\"unken, A. Potapov and S. Schlemmer,  Science
  \textbf{347} (6228), 1346--1349 (2015).

\bibitem{asv09}
O. Asvany and S. Schlemmer,  Int. J. Mass Spectrom.  \textbf{279}, 147--155
  (2009).

\bibitem{gup23}
D. Gupta, W.G.D.P. Silva, J.L. Doménech, E. Plaar, S. Thorwirth, S. Schlemmer
  and O. Asvany,  Faraday Discuss.  pp.~-- (2023),
  http://dx.doi.org/10.1039/D3FD00068K.

\bibitem{asv23}
O. Asvany, S. Thorwirth, P.C. Schmid, T. Salomon and S. Schlemmer,  Phys. Chem.
  Chem. Phys.  pp.~-- (2023), http://dx.doi.org/10.1039/D3CP01976D.

\bibitem{bru19}
S. Brünken, F. Lipparini, A. Stoffels, P. Jusko, B. Redlich, J. Gauss and S.
  Schlemmer,  J. Phys. Chem. A  \textbf{123} (37), 8053--8062 (2019), PMID:
  31422660.

\end{thebibliography}

\end{document}


\section{Linelist with comparison to former experiment}

Comparison between the transition frequencies (in wavenumbers) from the LOS 
measurements with those reported by Jagod et al.\ (J. Chem. Phys. 97, 7111–7123, 
1992).
For the transitions reported below, the $\Lambda$-doubling components (e and f) were separated with LOS for most of the transitions, while in the previous measurements from the Oka group, the splits in any of these lines could not be resolved. The lines for which the $\Lambda$-doubling splitting was not resolved are represented by a single frequency value.
The lines which were excluded from the upper state fit are marked with an asterisk.

\begin{verbatim}                   
P-branch               LOS 2023	    Oka
3 3.5 F1f 4 4.5 F1f    3126.4916    3126.493 
3 3.5 F1e 4 4.5 F1e    3126.4930              

3 2.5 F2e 4 3.5 F2e    3127.7120    3127.711 
3 2.5 F2f 4 3.5 F2f    3127.7147              

2 2.5 F1f 3 3.5 F1f    3128.6739    3128.675 
2 2.5 F1e 3 3.5 F1e                 

2 1.5 F2e 3 2.5 F2e    3130.0213    3130.023 
2 1.5 F2f 3 2.5 F2f    3130.0231              

1 1.5 F1f 2 2.5 F1f    3130.8393    3130.838  
1 1.5 F1e 2 2.5 F1e         

1 0.5 F2e 2 1.5 F2e    3132.3215    3132.321      
1 0.5 F2f 2 1.5 F2f         


Q-branch                                        
4 3.5 F2f 4 3.5 F2e    3135.6489 *              
4 3.5 F2e 4 3.5 F2f    3135.6617 *   

3 2.5 F2f 3 2.5 F2e    3135.6945              
3 2.5 F2e 3 2.5 F2f    3135.7027   

2 1.5 F2f 2 1.5 F2e    3135.7254    3135.722
2 1.5 F2e 2 1.5 F2f    3135.7300              

1 0.5 F2f 1 0.5 F2e    3135.7450    3135.747          
1 0.5 F2e 1 0.5 F2f    3135.7471     

6 6.5 F1e 6 6.5 F1f    3135.9362              
6 6.5 F1f 6 6.5 F1e    3135.9505       

5 5.5 F1e 5 5.5 F1f    3136.0069    3136.011         
5 5.5 F1f 5 5.5 F1e    3136.0164     

4 4.5 F1e 4 4.5 F1f    3136.0658    3136.068 
4 4.5 F1f 4 4.5 F1e    3136.0712              

3 3.5 F1e 3 3.5 F1f    3136.1140    3136.119           
3 3.5 F1f 3 3.5 F1e    3136.1166    

2 2.5 F1e 2 2.5 F1f    3136.1527    3136.154           
2 2.5 F1f 2 2.5 F1e    3136.1538    

1 1.5 F1e 1 1.5 F1f    3136.1804    3136.182 
1 1.5 F1f 1 1.5 F1e                  					           

R-Branch                                        
2 1.5 F2f 1 0.5 F2f    3139.1523    3139.150
2 1.5 F2e 1 0.5 F2e     
              
3 2.5 F2f 2 1.5 F2f    3141.4041    3141.402 
3 2.5 F2e 2 1.5 F2e                 
    
2 2.5 F1e 1 1.5 F1e    3141.4942    3141.491            
2 2.5 F1f 1 1.5 F1f 

3 3.5 F1e 2 2.5 F1e    3143.5970    3143.594 
3 3.5 F1f 2 2.5 F1f         
       
4 3.5 F2f 3 2.5 F2f    3143.6395 *  3143.636 
4 3.5 F2e 3 2.5 F2e    3143.6408 *              

4 4.5 F1e 3 3.5 F1e    3145.6913    3145.690
4 4.5 F1f 3 3.5 F1f    3145.6925     
          
5 4.5 F2f 4 3.5 F2f    3145.8598 *  3145.860 
5 4.5 F2e 4 3.5 F2e    3145.8610 *              

5 5.5 F1e 4 4.5 F1e    3147.7785    3147.781 
5 5.5 F1f 4 4.5 F1f    3147.7802              

6 5.5 F2f 5 4.5 F2f    3148.0640 *  3148.063
6 5.5 F2e 5 4.5 F2e    3148.0660 *

6 6.5 F1e 5 5.5 F1e    3149.8617    3149.862 
6 6.5 F1f 5 5.5 F1f    3149.8642     			

7 7.5 F1e 6 6.5 F1e    3151.9381    3151.942     
7 7.5 F1f 6 6.5 F1f    3151.9418    

8 8.5 F1e 7 7.5 F1e    3154.0114 *  3154.013 
8 8.5 F1e 7 7.5 F1f    3154.0164 *
\end{verbatim}

\section{Ground State Combination Difference Fit} 

The 68 ground state combination differences (in wavenumbers) used for the 
fit of the ground state are given below. The obs-calc rms is 0.0022~cm$^{-1}$.

\begin{verbatim}
Observed    Obs-Calc   v'  J'   N'           v"  J"   N"       
  3.4256    -0.0006    0  1.5   2    F2e  -  0  0.5   1    F2f 
  3.4223     0.0001    0  1.5   2    F2f  -  0  0.5   1    F2e 
  3.4269     0.0008    0  1.5   2    F2e  -  0  0.5   1    F2f 
  5.7023     0.0005    0  2.5   3    F2f  -  0  1.5   2    F2e 
  5.3411    -0.0001    0  2.5   2    F1f  -  0  1.5   1    F1e 
  3.4225     0.0003    0  1.5   2    F2f  -  0  0.5   1    F2e 
  5.7087    -0.0001    0  2.5   3    F2e  -  0  1.5   2    F2f 
  5.3402    -0.0003    0  2.5   2    F1e  -  0  1.5   1    F1f 
  5.7014    -0.0003    0  2.5   3    F2f  -  0  1.5   2    F2e 
  7.9907     0.0014    0  3.5   4    F2e  -  0  2.5   3    F2f 
  5.3415     0.0003    0  2.5   2    F1f  -  0  1.5   1    F1e 
  7.4788    -0.0007    0  3.5   3    F1e  -  0  2.5   2    F1f 
  5.7094     0.0006    0  2.5   3    F2e  -  0  1.5   2    F2f 
  7.9798     0.0012    0  3.5   4    F2f  -  0  2.5   3    F2e 
  5.3407     0.0001    0  2.5   2    F1e  -  0  1.5   1    F1f 
  7.4803    -0.0009    0  3.5   3    F1f  -  0  2.5   2    F1e 
  7.9906     0.0013    0  3.5   4    F2e  -  0  2.5   3    F2f 
 10.2532     0.0013    0  4.5   5    F2f  -  0  3.5   4    F2e 
  7.4804     0.0009    0  3.5   3    F1e  -  0  2.5   2    F1f 
  9.6250     0.0004    0  4.5   4    F1f  -  0  3.5   3    F1e 
  7.9791     0.0005    0  3.5   4    F2f  -  0  2.5   3    F2e 
  7.4825     0.0013    0  3.5   3    F1f  -  0  2.5   2    F1e 
  9.6210     0.0000    0  4.5   4    F1e  -  0  3.5   3    F1f 
 22.7933    -0.0001    0  5.5   6    F2e  -  0  3.5   4    F2e 
  9.6255     0.0009    0  4.5   4    F1f  -  0  3.5   3    F1e 
 11.7651    -0.0003    0  5.5   5    F1e  -  0  4.5   4    F1f 
  9.6213     0.0003    0  4.5   4    F1e  -  0  3.5   3    F1f 
 27.3303     0.0036    0  6.5   7    F2f  -  0  4.5   5    F2f 
 11.7638    -0.0016    0  5.5   5    F1e  -  0  4.5   4    F1f 
 13.9187    -0.0047    0  6.5   6    F1f  -  0  5.5   5    F1e 
 11.7716    -0.0002    0  5.5   5    F1f  -  0  4.5   4    F1e 
 13.9255     0.0021    0  6.5   6    F1f  -  0  5.5   5    F1e 
 16.0635    -0.0004    0  7.5   7    F1e  -  0  6.5   6    F1f 
 13.9137     0.0007    0  6.5   6    F1e  -  0  5.5   5    F1f 
 34.2991    -0.0051    0  8.5   8    F1f  -  0  6.5   6    F1f 
 40.8960    -0.0042    0  9.5   10   F2e  -  0  7.5   8    F2e 
 38.6147    -0.0011    0  9.5   9    F1e  -  0  7.5   7    F1e 
 45.3950     0.0024    0  10.5  11   F2f  -  0  8.5   9    F2f 
 42.9510     0.0001    0  10.5  10   F1f  -  0  8.5   8    F1f 
 49.8970    -0.0033    0  11.5  12   F2e  -  0  9.5   10   F2e 
 54.3680    -0.0049    0  12.5  13   F2f  -  0  10.5  11   F2f 
 64.6670     0.0015    0  15.5  15   F1e  -  0  13.5  13   F1e 
 72.2410    -0.0023    0  16.5  17   F2f  -  0  14.5  15   F2f 
 72.2600    -0.0041    0  16.5  17   F2e  -  0  14.5  15   F2e 
 76.7180     0.0009    0  17.5  18   F2e  -  0  15.5  16   F2e 
 73.3890     0.0008    0  17.5  17   F1e  -  0  15.5  15   F1e 
 76.6960     0.0012    0  17.5  18   F2f  -  0  15.5  16   F2f 
 81.1390    -0.0017    0  18.5  19   F2f  -  0  16.5  17   F2f 
 77.7710    -0.0045    0  18.5  18   F1f  -  0  16.5  16   F1f 
 81.1630    -0.0015    0  18.5  19   F2e  -  0  16.5  17   F2e 
 85.6070     0.0004    0  19.5  20   F2e  -  0  17.5  18   F2e 
 82.1250     0.0021    0  19.5  19   F1e  -  0  17.5  17   F1e 
 85.5800    -0.0014    0  19.5  20   F2f  -  0  17.5  18   F2f 
 90.0200     0.0028    0  20.5  21   F2f  -  0  18.5  19   F2f 
 90.0410    -0.0028    0  20.5  21   F2e  -  0  18.5  19   F2e 
 94.4760    -0.0004    0  21.5  22   F2e  -  0  19.5  20   F2e 
 94.4450    -0.0034    0  21.5  22   F2f  -  0  19.5  20   F2f 
 90.8890    -0.0022    0  21.5  21   F1f  -  0  19.5  19   F1f 
 98.8740    -0.0011    0  22.5  23   F2f  -  0  20.5  21   F2f 
 95.2670     0.0010    0  22.5  22   F1f  -  0  20.5  20   F1f 
 98.9070     0.0024    0  22.5  23   F2e  -  0  20.5  21   F2e 
103.3290     0.0004    0  23.5  24   F2e  -  0  21.5  22   F2e 
 99.6180     0.0048    0  23.5  23   F1e  -  0  21.5  21   F1e 
103.2950    -0.0027    0  23.5  24   F2f  -  0  21.5  22   F2f 
 99.6460     0.0045    0  23.5  23   F1f  -  0  21.5  21   F1f 
104.0190     0.0015    0  24.5  24   F1f  -  0  22.5  22   F1f 
103.9890     0.0011    0  24.5  24   F1e  -  0  22.5  22   F1e 
108.3620    -0.0007    0  25.5  25   F1e  -  0  23.5  23   F1e 

 
\end{verbatim}

\section{Fit of $\nu_3$}

Transition frequencies and fit residuals (in wavenumbers) for the assigned lines
in the $\nu_3$ vibrationally excited state (C-H asymmetric stretch).  

\begin{verbatim}
 Observed   Obs-Calc  v'  J'  N'          v"  J"  N"
3126.4916   -0.0016   1  3.5  3   F1f  -  0  4.5  4   F1f
3126.4930   -0.0013   1  3.5  3   F1e  -  0  4.5  4   F1e
3127.7120   -0.0052   1  2.5  3   F2e  -  0  3.5  4   F2e
3127.7147   -0.0044   1  2.5  3   F2f  -  0  3.5  4   F2f
3128.6735    0.0009   1  2.5  2   F1f  -  0  3.5  3   F1f
3128.6739    0.0006   1  2.5  2   F1e  -  0  3.5  3   F1e
3130.0213    0.0004   1  1.5  2   F2e  -  0  2.5  3   F2e
3130.0231    0.0007   1  1.5  2   F2f  -  0  2.5  3   F2f
3130.8393    0.0007   1  1.5  1   F1f  -  0  2.5  2   F1f
3130.8395    0.0005   1  1.5  1   F1e  -  0  2.5  2   F1e
3132.3215    0.0046   1  0.5  1   F2e  -  0  1.5  2   F2e
3132.3227    0.0044   1  0.5  1   F2f  -  0  1.5  2   F2f
3135.6945   -0.0032   1  2.5  3   F2f  -  0  2.5  3   F2e
3135.7027   -0.0038   1  2.5  3   F2e  -  0  2.5  3   F2f
3135.7254    0.0012   1  1.5  2   F2f  -  0  1.5  2   F2e
3135.7300    0.0004   1  1.5  2   F2e  -  0  1.5  2   F2f
3135.7450    0.0046   1  0.5  1   F2f  -  0  0.5  1   F2e
3135.7471    0.0040   1  0.5  1   F2e  -  0  0.5  1   F2f
3135.9362   -0.0010   1  6.5  6   F1e  -  0  6.5  6   F1f
3135.9505    0.0002   1  6.5  6   F1f  -  0  6.5  6   F1e
3136.0069   -0.0006   1  5.5  5   F1e  -  0  5.5  5   F1f
3136.0164    0.0005   1  5.5  5   F1f  -  0  5.5  5   F1e
3136.0658   -0.0010   1  4.5  4   F1e  -  0  4.5  4   F1f
3136.0712   -0.0005   1  4.5  4   F1f  -  0  4.5  4   F1e
3136.1140   -0.0013   1  3.5  3   F1e  -  0  3.5  3   F1f
3136.1166   -0.0012   1  3.5  3   F1f  -  0  3.5  3   F1e
3136.1527   -0.0001   1  2.5  2   F1e  -  0  2.5  2   F1f
3136.1538   -0.0000   1  2.5  2   F1f  -  0  2.5  2   F1e
3136.1798    0.0002   1  1.5  1   F1e  -  0  1.5  1   F1f
3136.1804    0.0005   1  1.5  1   F1f  -  0  1.5  1   F1e
3139.1523    0.0020   1  1.5  2   F2f  -  0  0.5  1   F2f
3139.1525    0.0007   1  1.5  2   F2e  -  0  0.5  1   F2e
3141.4039   -0.0026   1  2.5  3   F2f  -  0  1.5  2   F2f
3141.4041   -0.0041   1  2.5  3   F2e  -  0  1.5  2   F2e
3141.4942    0.0002   1  2.5  2   F1e  -  0  1.5  1   F1e
3141.4945    0.0001   1  2.5  2   F1f  -  0  1.5  1   F1f
3143.5965    0.0000   1  3.5  3   F1e  -  0  2.5  2   F1e
3143.5970   -0.0003   1  3.5  3   F1f  -  0  2.5  2   F1f
3145.6913   -0.0000   1  4.5  4   F1e  -  0  3.5  3   F1e
3145.6925   -0.0002   1  4.5  4   F1f  -  0  3.5  3   F1f
3147.7785   -0.0008   1  5.5  5   F1e  -  0  4.5  4   F1e
3147.7802   -0.0011   1  5.5  5   F1f  -  0  4.5  4   F1f
3149.8617    0.0011   1  6.5  6   F1e  -  0  5.5  5   F1e
3149.8642    0.0008   1  6.5  6   F1f  -  0  5.5  5   F1f
3151.9381    0.0024   1  7.5  7   F1e  -  0  6.5  6   F1e
3151.9418    0.0026   1  7.5  7   F1f  -  0  6.5  6   F1f


\end{verbatim}